\documentclass[a4paper,12pt]{article}
\pdfoutput=1 

\usepackage{jheppub} 
\usepackage{subfigure}
\usepackage{hyperref}
\definecolor{refkey}{gray}{0.45}
\definecolor{labelkey}{RGB}{155,48,48}
\usepackage{float}
\usepackage{mathtools}
\usepackage{physics}
\usepackage{simplewick}
\usepackage[obeyFinal]{todonotes}
\usepackage[normalem]{ulem}
\usepackage {cancel}

\def\beq{\begin{eqnarray}}\def\eeq{\end{eqnarray}}
\def\be{\begin{equation}}\def\ee{\end{equation}}

\def\mes[#1]{d^{3}{#1}}

\def\del{\partial}

\newcommand{\half}{\frac{1}{2}}

\def\del{\partial}

\def\order{\ensuremath{\mathcal{O}}}

\reversemarginpar

\definecolor{UI_blue}{RGB}{32, 64, 151}
\definecolor{UI_red}{RGB}{187, 62, 24}
\definecolor{UI_blue2}{RGB}{0, 84, 147}
\definecolor{UI_red2}{RGB}{159, 32, 66}
\definecolor{UI_gray}{RGB}{169, 169, 169}
\definecolor{UI_sepia}{RGB}{112, 66, 20}
\definecolor{UI_bittersweet}{RGB}{254, 111, 94}
\definecolor{UI_emerald}{RGB}{80, 200, 120}
\definecolor{UI_olivegreen}{RGB}{181, 179, 92}
\definecolor{UI_cadetblue}{RGB}{95, 158, 160}
\definecolor{UI_fuchsia}{RGB}{255, 0, 255}
\definecolor{UI_midnightblue}{RGB}{25, 25, 112}
\definecolor{UI_royalblue}{RGB}{0,35, 102}
\definecolor{UI_periwinkle}{RGB}{204, 204, 255}
\definecolor{UI_redorange}{RGB}{255, 83, 73}
\definecolor{UI_brickred}{RGB}{203,65,84}	
\definecolor{UI_forestgreen}{RGB}{34, 139, 34}
\definecolor{UI_tan}{RGB}{210,180,140}	
\definecolor{UI_burlywood}{RGB}{222,184,135}
\definecolor{UI_burlywood}{RGB}{192,64,0}
\definecolor{UI_darkorchid}{RGB}{153,50,204}

\usepackage{titlesec}
\usepackage{verbatim}
\usepackage{fancyhdr}
\usepackage{enumerate}
\hypersetup{colorlinks=true,citecolor=blue,urlcolor=black}
\usepackage{slashed, enumitem}
\usepackage{array}
\newcolumntype{P}[1]{>{\centering\arraybackslash}p{#1}}
\usepackage{anyfontsize}
	\vspace{-11cm}
	
	\author[a, b]{Norihiro Iizuka}
	\author[b]{and Sunil Kumar Sake}

	\affiliation[a]{\it Department of Physics, National Tsing Hua University, Hsinchu 30013, Taiwan}
	\affiliation[b]{\it Yukawa Institute for Theoretical Physics, Kyoto University, Kyoto 606-8502, Japan}

	\emailAdd{iizuka@phys.nthu.edu.tw}
	\emailAdd{sunilsake1@gmail.com}
	
	\vspace{1cm}

	\abstract{ We explore a Centaur geometry in JT gravity, which is an asymptotically AdS spacetime but in the IR admits a dS bubble with another AdS geometry in the deep IR.  Thus, this geometry admits a holographic dual in the sense that it is  asymptotically AdS. In an attempt to understand this geometry, we calculate the density of states of the putative boundary dual for such mixed geometries by evaluating the on-shell action. We compute the density of states analytically in the classical limit. The resultant density of states suggest that the degrees of freedom in the IR are reduced in such a putative boundary theory due to the IR modification corresponding to the dS bubble. }

\title{A note on Centaur geometry -  probing IR de Sitter spacetime holography}

\preprint{\parbox{3cm}{YITP-24-143}}
\begin{document}
	\maketitle
	\flushbottom
	
	\vskip 10pt

	\section{Introduction and motivation}
	\label{introd}
	Even though the discovery of holography for asymptotically Anti-de Sitter (AdS) spacetime has greatly enhanced our understanding of theories of gravity \cite{Maldacena:1997re}, we are far from such an understanding for more realistic spacetime, the de Sitter (dS) spacetimes, which corresponds to the current universe we live in.
 	The problem of understanding the holography for dS has been a long-standing one. Although string theory constructions of dS have been proposed since \cite{Kachru:2003aw}, the nature of the holographic dual for the de Sitter, if it exists, is very obscure.  Some of the approaches involve taking the holographic boundary at the future infinity of dS  \cite{Strominger:2001pn, Bousso:2001mw,Balasubramanian:2001nb}. This route of finding a holographic dual for dS at the future boundary leads to a CFT that has peculiar properties, an imaginary central charge being one of them. Recent literature has seen a rise in alternate approaches to this problem such as cosmological bootstrap (see \cite{Baumann:2022jpr} and the references therein), stretched horizon holography \cite{Susskind:2021esx, Susskind:2021omt} and static patch holography {\cite{Anninos:2011af,Leuven:2018ejp}}. In the approaches of static patch holography and stretched horizon holography, a time-like boundary in the static patch of the dS is considered to be the holographic boundary. The dual CFT is then assumed to live on this boundary. This seems a bit weird as the gravity is dynamic in the static patch region and so it seems artificial to place a boundary in a region where fluctuations of gravity are important. 
 	
 	One possible way to get around this issue is to embed the dS region in a larger AdS spacetime. The artificial boundary then becomes the location of the domain wall between the AdS and dS spacetimes. This approach has the advantage that the holographic dual lives at the timelike boundary of AdS. Given our understanding of the holographic dual to asymptotically AdS spacetimes, we can hope to leverage this to understand the same for the dS spacetimes in such settings. However in a generic spacetime dimension where $d >2$, there are obstructs, the dS spacetimes are always hidden behind the AdS black hole horizons \cite{Freivogel:2005qh}. However in two-dimensional spacetime, one can avoid this issue. These kinds of two-dimensional geometries, dubbed Centaur geometries, were proposed first in \cite{Anninos:2017hhn} for a particular class of dilaton gravity theories, called the Jackiw-Teitelboim (JT) gravity, and since then studied actively in the recent past, see for example, \cite{Anninos:2017cnw, Anninos:2018svg, Chapman:2021eyy,Anninos:2022hqo,Anninos:2017hhn}. {Thus it is especially easier to study dS from holographic perspective in a two-dimensional setting.  In this paper, we focus on the JT gravity. In JT gravity for pure dS, one can directly study the Schwarzian theory on the spacelike boundary at $t = \infty$ \cite{Maldacena:2019cbz}. We shall consider a more general potential to embed dS region in AdS spacetime.}
	
	The action for this theory is given by 
 	\begin{align}
 		I_{\text{JT,bulk}}=- \half \int d^2x \sqrt{-g}(\phi R+U(\phi))\label{dilgr}
 	\end{align}
 	The case of $U=2\phi$ corresponds to the AdS JT gravity as can be noticed easily from the fact that the dilaton equations of motion lead to $R=-2$. Similarly, the case of $U=-2\phi$ corresponds to the dS case. Since we are interested in the geometries that interpolate between AdS and dS, we should consider $U(\phi)$ that interpolates between that for AdS and dS. These kinds of Centaur geometries cannot be understood as perturbative corrections to the AdS geometry.

	 In this paper, we are interested in computing the density of states of the putative boundary dual of the bulk Centaur geometry. To compute this density of states, we would have to obtain the partition function at finite temperature\footnote{This corresponds to evaluating the Euclidean path integral with appropriate initial and final boundary condition. The fixed temperature criterion corresponds to computing the path integral with a fixed renormalized length of the boundary as the final boundary condition. The natural initial boundary condition from the path integral perspective is the analog of no-boundary boundary condition. }. However, it is not easy to compute the path integral beyond the on-shell  approximation. So, we shall discuss the computation of the density of states in the on-shell approximation and highlight some features. Let us also  mention that a similar study of more general spacetimes that asymptote to dS potential at large values of the dilaton field, rather than the AdS potential considered in this paper, is carried out in \cite{kanhucentaur1}. The ideas related to the computation of the density of states in the on-shell approximation in this paper are similar to those considered in \cite{kanhucentaur1}.
	 
	 On the other hand, the canonical quantization of JT gravity with a general potential can be carried out straightforwardly.  The procedure of canonical quantization involves solving the Wheeler-DeWitt (WDW) equation  \cite{DeWitt:1967yk}. For JT theory, this equation can be solved exactly and the general solutions for the JT gravity, with the appropriate boundary conditions, are found \cite{Hennauxjt}. Since the WDW equation is a differential equation, appropriate boundary conditions should be specified to uniquely identify a  solution\footnote{In addition, there is uncertainty regarding the order of operators when turning the classical Einstein equations into quantum WDW equations for operators. This uncertainty corresponds to the uncertainty in how the contour is defined in the path integrals.}. Usually, this boundary condition is taken to be specified by an appropriate path integral\footnote{For example, for JT gravity, the path integral can be done for asymptotic boundary conditions 	 \be
	 \phi \rightarrow \infty, \quad \ell \rightarrow \infty, \quad \frac{\phi}{\ell}=\text{const}
	 \ee 
	 where $\ell$ is the length of the boundary. This is taken as the boundary condition that fixes the solution to the WDW equation uniquely. The unique solution thus obtained gives an expression that is valid for all $\phi, \ell$. This can be checked against the path integral result in cases where the path integral can be done.}. 	
The exact specification of the state depends on a one-parameter function which is usually fixed by the requirement of the matching with the path integral answer. 
 	 
	This paper is structured as follows. In section 2, we summarize the classical solutions of the Centaur geometries and discuss some thermodynamical aspects related to the stability of various solutions.  In section 3 we review the WDW equation and its solution for general potentials. In section 4, we outline the computation of the density of states for a general potential, in the WKB limit. Following this, in section 5,  we specialize to the particular case of Centaur geometry and obtain the corresponding density of states. We end with some conclusions in section 6.

	\section{Classical Centaur geometries in JT gravity}
	\label{rev}
	In this section, we review and summarize classical Centaur solutions in JT gravity.

The Euclidean action for the general dilaton gravity model in the class of JT gravity theories is given by 
	\begin{align}
		I_{\text{JT,bulk}}=-\frac{1}{2}\int (\phi R+U) \,. \label{eucjt}
	\end{align}
Note that we are working in units where $8\pi G_N=1$, where $G_N$ is the two-dimensional Newton's constant. In addition to the bulk term in eq.~\eqref{eucjt}, there will be also be boundary terms, the conventional Gibbons-Hawking-York term, that is necessary for a well-defined variational principle {with Dirichlet boundary conditions for the metric and the dilaton, $\phi$}. Also, counter-terms may also be required. For now let us just add the Gibbons-Hawking-York term given by 
\begin{align}
	I_{\text{GH}}=-\int_\del \phi K \,. \label{gh}
\end{align}
The full JT action is then given by 
\begin{align}
	I_{\text{JT}}=I_{\text{JT,bulk}}+I_{\text{GH}} \,. \label{fulljtac}
\end{align}


Let us begin by presenting the on-shell solutions for the JT gravity system. The equations of motion obtained by varying the dilaton and the metric are given by 
\begin{align}
	R+U'(\phi)=0 \,, \quad 
	\nabla_\mu\nabla_\nu\phi-g_{\mu\nu}\nabla^2\phi+\half g_{\mu\nu}U =0 \,. \label{eom}
\end{align}
Taking the dilaton as one of the coordinates, we can write the Euclidean solutions as
\begin{align}
		\phi=r \,, \quad 
		ds^2=f(r)d{x}^2+\frac{dr^2}{f(r)} \,,\quad f(r)=\int^r_{r_h} d\phi\,\, U(\phi) \,.\label{onsol}
\end{align}
where $r_h$ is the location of the horizon, where $f(r_h)=0$. Note that in two-dimensional spacetime $r_h$ can be negative. 

Of particular interest for us will be the geometries which have AdS in the UV and interpolate to dS spacetime in the IR. We shall call the geometries corresponding to such mixed potentials as Centaur geometries, following the terminology in the literature \cite{Anninos:2017hhn}.

The case of AdS and dS correspond to the potential \cite{Anninos:2018svg,Anninos:2022hqo}
\begin{align}
	U_{\rm AdSdS}(\phi)=\begin{cases}
		2\phi \qquad \text{AdS}\\
		-2\phi\qquad \text{dS}
	\end{cases}
	\label{Uphi}
\end{align}
The above potential $U$ that interpolates between AdS and dS, written compactly reads,
\begin{align}
	U_{\rm AdSdS}(\phi)=2\phi\,\,\text{sign}(\phi) \,. \label{uphiforce}
\end{align}

Although the potential is smooth, its derivatives are discontinuous at the location of the domain wall, $\phi=0$, in the present case. However we can always make potential smooth. One such potential is given by 
\begin{align}
	U_{\rm AdSdS}(\phi)&=2\phi\tanh(\frac{\phi}{\epsilon})  , \qquad  (\epsilon \ll 1) \label{uphtan}  \\
&=\begin{cases}
		2\phi  \qquad \text{AdS} \qquad  \phi \gg \epsilon \,,\\
		-2\phi \qquad \text{dS}  \qquad  \phi \ll -\epsilon \,,
		\label{uphtanasym} 
	\end{cases}
\end{align}
This potential has following asymptotics. The $\tanh$ function smoothens out the cusp near $\phi=0$. This potential differs from the one in eq.~\eqref{uphiforce} in a region $\abs{\phi}\sim \epsilon \ll 1$, thus by taking $\epsilon$ to be arbitrarily small, it can be made as close to eq.~\eqref{uphiforce} as we want.  

As we will see, the configuration with dS horizon is thermodynamically unstable  \cite{Witten:2020ert}.  On the other hand, the configuration of the horizon in the AdS region is stable. Since we are interested in exploring the properties due to the dS region, we would want the geometry to include observable dS region from the AdS boundary. 
Let us review this argument now. 

If there is a black hole with $r = r_h$, then $U(r_h) > 0$, where otherwise, the metric function $f(r)$ eq.~\eqref{onsol} becomes negative near $r_h$ where $r > r_h$, 
\begin{align}
 f(r)=\int^r_{r_h} d\phi\,\, U(\phi)  = U(r_h) (r-r_h) + \order(r-r_h)^2 \, \quad \mbox{(at $r \gtrsim r_h$)}
\end{align}
Thus $U(r_h)$ is related to the black hole temperature, 
\begin{align}
T = \frac{U(r_h)}{4 \pi}
\label{BHtemperature}
\end{align}

The entropy is given by the value of the dilaton at the horizon, upto a factor of $2\pi$, 
\begin{align}
S = 2 \pi r_h
\end{align}

Since we restrict to asymptotically AdS$_2$, $U(\phi) \sim 2 \phi$ for $\phi \to \infty$, thus we can identify the energy from the asymptotic form of $f(r)$ as   
\begin{align}
\label{defofM}
 f(r)=  r^2  - 2 E \, \quad \mbox{(at $r \to \infty$)}
\end{align}
This $E$ can be seen as ``energy'' from the consistency of the thermodynamical law as  
\begin{align}
2  \frac{dE}{dr_h} = {U(r_h)}  =  4 \pi T =  2   T   \frac{dS}{dr_h} \,.
\end{align}

Furthermore, positive specific heat implies $U'(r_h) > 0$, since 
\begin{align}
C\equiv\frac{dE}{dT} \propto  {\frac{dE}{dr_h}}/{\frac{dT}{d r_h}}  \propto \frac{U(r_h)}{U'(r_h)} > 0  \,. 
\end{align}
Thus, the horizon in dS has negative specific heat since there $U'(r_h) < 0$. 

We would like to focus on the Centaur geometry where at large $r \to \infty$, we have AdS geometry. However for the potential in eq.~\eqref{uphtan} the specific heat for the dS horizon is negative. Therefore, the dS horizon will be unstable and will collapse to an AdS black hole. Thus, from the outside, the dS region is inaccessible because of the AdS horizon, which is not interesting. 

\begin{figure}[h!]
	\centering

	\tikzset{every picture/.style={line width=0.75pt}} 
	
	\begin{tikzpicture}[x=0.75pt,y=0.75pt,yscale=-1,xscale=1]
		
		\draw  (166.5,215.75) -- (473.5,215.75)(197.2,39.35) -- (197.2,235.35) (466.5,210.75) -- (473.5,215.75) -- (466.5,220.75) (192.2,46.35) -- (197.2,39.35) -- (202.2,46.35)  ;
		\draw  [dash pattern={on 4.5pt off 4.5pt}]  (197.2,215.75) -- (155,291) ;
		\draw    (250,114) -- (195.2,220.5) ;
		\draw    (250,114) .. controls (272,73) and (281,47) .. (319,108) ;
		\draw    (319,108) .. controls (341,140) and (367,112) .. (402,92) ;
		\draw    (402,92) -- (459,56) ;
		\draw  [dash pattern={on 4.5pt off 4.5pt}]  (283.2,72.75) -- (280,215) ;
		\draw  [dash pattern={on 4.5pt off 4.5pt}]  (343.2,122.75) -- (342,215) ;
		\draw    (196,97) -- (462,98) ;
		\draw  [dash pattern={on 4.5pt off 4.5pt}]  (364,115) -- (116,263) ;
		
		\draw (211,128.84) node [anchor=north west][inner sep=0.75pt]    {$c\phi $};
		\draw (164,15.4) node [anchor=north west][inner sep=0.75pt]    {$U( \phi )$};
		\draw (468,89.4) node [anchor=north west][inner sep=0.75pt]    {$4\pi T	$};
		\draw (430,41.84) node [anchor=north west][inner sep=0.75pt]    {$2\phi $};
		\draw (272,217.4) node [anchor=north west][inner sep=0.75pt]    {$\phi _{1}$};
		\draw (335,217.4) node [anchor=north west][inner sep=0.75pt]    {$\phi _{2}$};
		\draw (482,206.4) node [anchor=north west][inner sep=0.75pt]    {$\phi $};
		
	\end{tikzpicture}
			\caption{The potential $U(\phi)$ we consider in this paper. This admits Centaur geometry. Here $c > 2$.}
	\label{fig31}
\end{figure}
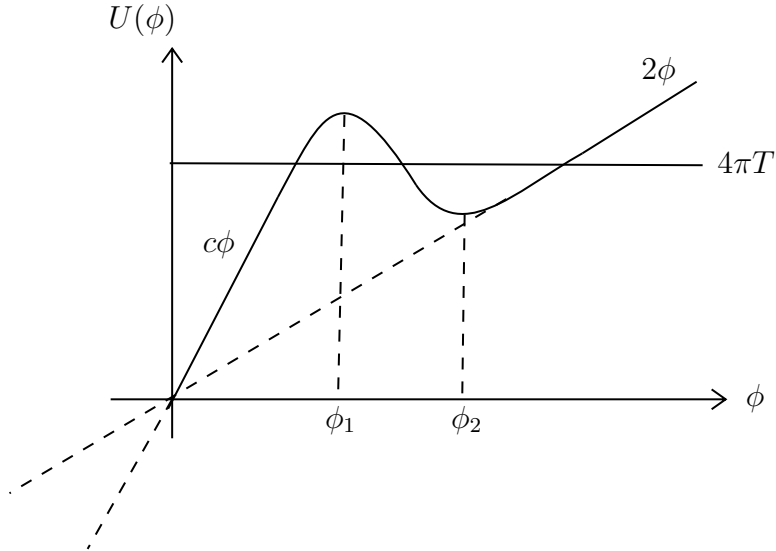

Thus, we shall consider the potential of the qualitative form as Fig.~\ref{fig31} and it can be modeled crudely as below \cite{Anninos:2022hqo}, 
\begin{align}
	U(\phi)=\begin{cases}
		c\,\phi \qquad &\phi<\phi_1    , \qquad (c > 2) \\
		U_0-\alpha\,\phi\qquad& \phi_1<\phi<\phi_2\\
		2\,\phi \qquad &\phi_2<\phi
	\end{cases}
\label{qsimp}
\end{align}

For $\phi>\phi_2$, the potential is exactly the same as the pure AdS case. The potential is different only in the IR region, for $\phi<\phi_2$. In this region, we have a dS spacetime.  For even smaller values of $\phi$ such that $\phi<\phi_1$, the spacetime is again AdS but with a different cosmological constant. Now, in this setup, for a fixed temperature, there could be either one horizon or three horizons. The location of the horizon for a fixed temperature is determined by the condition \eqref{BHtemperature}. 

For temperature such that $ 4 \pi T<U(\phi_2)$ or $ 4 \pi T > U(\phi_1)$, there is a single horizon. For temperature in the intermediate region such that $U(\phi_2)< 4 \pi T<U(\phi_1)$, the are three possible horizons, one in the dS region and two in the AdS regions. However, the dS horizon, as mentioned earlier is unstable. The two AdS horizons are stable but the dominant one depends on the exact value of the temperature and the lower free energy configuration dominates. As we shall show in detail now, for $T<T_c$ the horizon in the inner AdS dominates and for $T>T_c$, the horizon in the outer AdS dominates, { where $T_c$, called transition temperature which depends on the parameters of the potenial, delineates the two possibliites mentioned above.} For $T<T_c$, the configuration with the inner horizon is the dominant one and it has an dS region outside the horizon. This is the configuration that we would be interested in analyzing. The actual value of the transition temperature $T_c$ depends on the exact details of the potential. For our specific potential in Fig.~\ref{fig31}, the transition temperature, $T_c$ are calculated later in section \ref{explicitwf}.

Continuity of the potential at $\phi=\phi_1$ and $\phi = \phi_2$ implies the following relations between various parameters
\begin{align}
	 \alpha=\frac{c\phi_1-2\phi_2}{\phi_2-\phi_1}\,, \quad 
 U_0 =  \frac{(c-2) \phi_1 \phi_2}{\phi_2 - \phi_1} \,.
	\label{relbpar}
\end{align}
Instead of $c$, $\alpha$, $U_0$, we regard $c$, $\phi_1$, $\phi_2$ as parameters of our potential. 

The qualitative nature of the potential that we would be interested in studying is as shown in Fig.~\ref{fig31}. 
{Even though \eqref{qsimp} has discontinuous derivative, one can make it smooth as Fig.~\ref{fig31}. 
For example, one can consider a potential of the form
\begin{align}
	U(\phi)=\phi\left(c_1+c_2\tanh(k_1\phi)+c_2\tanh(k_2\phi+\phi_c)\right)\label{smoothpot}.
\end{align} By appropriately tuning the parameters $c_1,c_2,c_3,k_1,k_2,\phi_c$, we can obtain a smoothed version of the potential in eq.\eqref{qsimp}. }

	\section{Canonical Quantization and WDW equation}
\label{canquz}
Let us now set up the procedure of canonical quantization of JT gravity. We shall carry this out for a general potential $U(\phi)$, not specializing to AdS or dS spacetime explicitly since our focus will be to consider the Centaur geometries that interpolate between AdS and dS. This part is meant to be a quick review of some of the known results that we intend to use later on. 
	Working in the ADM gauge with radial coordinate $r$ playing the role of time, 
	\begin{align}
		ds^2=N^2dr^2+g_1(dx+N_\perp dr)^2\label{dsadm}
	\end{align}
	the action is given by, after several integration by parts, 
	\begin{align}
		-I_{\text{JT}}=&\int d^2x \left[\frac{\dot{\phi}}{N}\left(-\frac{N_{\perp}}{2\sqrt{g_1}}{g_1'}-\sqrt{g_1} N_{\perp}'+\frac{\dot{g}_1}{2\sqrt{g_1}}\right)\right]\nonumber\\
		&+\int d^2x\left[ \frac{\phi'}{N}\left(\frac{NN'}{\sqrt{g_1}}+\sqrt{g_1}N_{\perp}N_{\perp}'-\frac{N_{\perp}}{2\sqrt{g_1}}\dot{g}_1+\frac{N_{\perp}^2}{2\sqrt{g_1}}g_1'\right)+\frac{1}{2}NU(\phi)\sqrt{g_1}\right]\label{jtadmact}
	\end{align}
	In the above, dots denote derivatives with respect to $r$ and primes denote derivatives with respect to $x$. 
	The canonical momenta are given by
	\begin{align}
		\pi_{N}=0 \,, \quad \pi_{N_{\perp}} &= 0 \,, \quad \pi_{g_1}=-\frac{\dot{\phi}}{2N\sqrt{g_1}}+\frac{N_{\perp}\phi'}{2N\sqrt{g_1}} \,, \nonumber\\
		\pi_\phi&=\frac{N_{\perp}}{2N\sqrt{g_1}}g_1'+\frac{N_{\perp}'\sqrt{g_1}}{N}-\frac{\dot{g}_1}{2N\sqrt{g_1}}	\,. \label{canmomall}
	\end{align}

	Also, the constraint equations, the equations of motion obtained by varying with respect to $N,N_{\perp}$, are, 
	\begin{align}
		0=\mathcal{H}\equiv{\frac{\delta 	I_{\text{JT}}}{\delta N}}&=-\left( \frac{\sqrt{g_1}U}{2}-\frac{1}{N^2}(\dot{\phi}-\phi'N_{\perp})\left(\frac{\dot{g}_1}{2\sqrt{g_1}}-\sqrt{g}_1 N_\perp' -\frac{N_\perp g_1'}{2\sqrt{g_1}}\right)-\left(\frac{\phi'}{\sqrt{g_1}}\right)' \right)\nonumber\\
		&=2\pi_\phi \pi_{g_1}\sqrt{g_1}+ \left(\frac{\phi'}{\sqrt{g_1}}\right)'-\half \sqrt{g_1}U\nonumber\\
		0=\mathcal{P}\equiv \frac{\delta I_{\text{JT}}}{\delta N_\perp}&=-\left( \left(\frac{\dot{\phi}\sqrt{g_1}}{N}\right)'-\frac{\dot{\phi}g_1'}{2N\sqrt{g_1}}+\frac{\phi' g_1' N_\perp}{N\sqrt{g_1}}+\frac{\phi'\sqrt{g_1}N_\perp'}{N}-\frac{\phi'\dot{g}_1}{2N\sqrt{g_1}}-\left(\frac{\phi '\sqrt{g_1}N_\perp}{N}\right)'\right)
		\nonumber\\
		&=2g_1\pi_{g_1}'+\pi_{g_1}g_1'-\pi_\phi \phi'\label{hammomcon}
	\end{align}
The Hamiltonian and momentum constraints have the usual operator ordering ambiguities and also coincident singularities. To overcome these, following \cite{Hennauxjt}, we first simplify the constraints classically and then impose them. 
	To eliminate $\pi_\phi$, we take a linear combination, 
	\begin{align}
		0=2\pi_{g_1}\sqrt{g_1}\mathcal{P}+\phi' \mathcal{H}=\sqrt{g_1}\pqty{2g_1\pi_{g_1}^2+\frac{1}{2}\left(\frac{\phi'}{\sqrt{g_1}}\right)^2-\half W(
			\phi)}' \,,
				\end{align}
					where 
	\begin{align}
		W(\phi)=\int^\phi_{0} U(x)dx  \,.
		\label{defofW}
	\end{align}
By integrating this equation with respect to {{$x$}}, we obtain
	\begin{align}
		g_1\pi_{g_1}^2+\frac{1}{4}\left(\frac{\phi'}{\sqrt{g_1}}\right)^2-\frac{1}{4}W(\phi)=-\frac{1}{4}M	 \,,
		\label{lcomcon}
	\end{align}
	where $M$ is an integration constant, which corresponds to half of the ADM mass of the state corresponding to \eqref{defofM}.
	Thus, we obtain the following equation, 
		\begin{align}
		\pi_{g_1} = \pm \frac{1}{2}\sqrt{ \left(  \frac{W-M}{ g_1}	 - \left(\frac{\phi'}{{g_1}}\right)^2 \right) }\,,
				\label{lcomcon2}
	\end{align}\\
	We are now in a position to quantize the theory. Note first that $r$ is an Euclidean coordinate, instead of Lorenzian time, thus the quanziation condition becomes\footnote{In ordinary quantum mechanics with conventional kinetic term, we have quantization condition $ \left[ x \,, p \right] = i $, where $p =  m \dot{x}$. By making the analytic continuation of time as $ t = -i \tau$, we have $p_E = m \frac{dx}{d\tau} =- i p$. Thus we have quantization condition $ \left[ x \,, p_E \right] = 1 $.}  
	\begin{align}
	\left[ g_1 \,, \pi_{g_1} \right] = 1 \, \quad \Rightarrow \quad \pi_{g_1} = - \frac{\partial}{\partial g_1}  \,,
	\end{align}
	thus, the WDW equation for a wave function $\Psi$ as 
	\begin{align}
		\pi_{g_1}\Psi=- \del_{g_1}\Psi=\pm \frac{1}{2 g_1} \left(\sqrt{ g_1 \left( {W(\phi)-M}\right)-\left({\phi'}\right)^2}\right)\Psi  \,.
		\label{pig1}
	\end{align}
We impose this as a constraint on the physical wave functions $\Psi$ in the quantum theory. 
	The solution for this is given by
		\begin{align}
				\Psi& =e^{\pm  S},
				\end{align}
				where
				\begin{align} S&= \int dx \left[ \sqrt{g_1(W(\phi) - M) - \,  \phi'^2} -\phi' \tan^{-1}\left(\frac{\sqrt{g_1(W(\phi) - M) - \,  \phi'^2}}{\phi'}\right)  \right] \,,   \label{finwfforw}
	\end{align}
	Let us choose a slice on which $\phi'=0$, {\it i.e.,} constant $\phi$ slice. Then the wave function at that slice becomes
	\begin{align}
		\Psi(\phi)=e^{\pm  \ell \sqrt{W(\phi)-M}} \,, \quad \mbox{where} \quad \ell = \int dx \sqrt{g_1} \,.\label{psioncdil}
	\end{align}
	Since this contains the integration constant $M$, 
	a more general solution would be 
	\begin{align}
		\Psi(\phi)=\int dM \left( \rho_+(M)e^{+  \ell \sqrt{W(\phi)-M}} + \rho_-(M)e^{-  \ell \sqrt{W(\phi)-M}}  \right)  \,.\label{rhois}
	\end{align}
	Different $\rho(M)$ correspond to different theories. One way to fix the choice of $\rho(M)$ is to compute the wave function from the path integral and match it in the asymptotic region of large $\ell$, $\phi$. 
	
	For example, doing this for Euclidean AdS$_2$ gives the density of states as
	\begin{align}
		\rho_+(M)=\sinh(2\pi\sqrt{M}) := \rho_0(M) \,,  \quad \rho_-(M)  = 0 \,. \label{rhodos}
	\end{align}
	Here for later purpose, we call the AdS$_2$ density of states as $\rho_0$. 
For Euclidean AdS$_2$ case we have 
	\begin{align}
	 U(\phi) = 2 \phi \,, \quad W(\phi) = \phi^2 \,, 
	 \end{align}
and thus the wave function becomes	
	 \begin{align}
	  \Psi(\phi)=\int dM \sinh(2\pi\sqrt{M}) e^{  \ell \sqrt{\phi^2-M}} \,, 
	\end{align}
		Note that 
	near the AdS boundary, we have $\phi \to \infty$, and 
	\begin{align}
	e^{ \ell \sqrt{\phi^2-M}}  \approx e^{\ell \phi  - \frac{\ell M}{2 \phi }}
	\end{align}
	The factor $e^{\ell \phi}$ is the divergent contribution which must be eliminated by the counter term. In fact, from the classical action eq.~\eqref{gh},  one obtains the boundary contribution as 
	\begin{align}
	I_{GH} = - \phi \int dx  K \, \quad \mbox{with} \quad K = 1 \, \quad \Rightarrow  \quad e^{-I_{GH}} = e^{\phi \ell} \,.
	\end{align}
	This yields the divergence which must be removed by the 
	counter term. Then the finite part of the wave function is the partition function. { Denoting} it as $Z_0(\ell, \phi)$ , it becomes 
		 \begin{align}
		 \label{zerothpartition}
	 Z_0(\phi, \ell) &= e^{-\ell \phi} \Psi(\phi) \approx \int_0^\infty dM \sinh(2\pi\sqrt{M}) e^{ - \frac{\ell M}{2 \phi }}\nonumber\\
	 &  \approx \int_0^\infty dM e^{2\pi\sqrt{M}- \frac{\ell M}{2 \phi }} \propto \left( {\frac{\phi}{\ell}} \right)^{3/2} e^{2 \pi^2 \frac{ \phi}{\ell}}\,, 
	\end{align}
	{The integral in the first line above, which is just a Laplace transform, can be exactly evaluated exactly. However, a saddle point approximation of the integral also gives the same result as the exact result. For centaur geometries mentioned earlier, we would only be able to do a saddle point approximation. The saddle point answer in the second line is a good approximation when
}
	\begin{align}
	M_{\rm saddle} \approx \left( 2 \pi \frac{ \phi}{\ell} \right)^2 \gg 1 \,\Rightarrow\frac{\phi}{l}\gg 1
	\label{Msaddle}
	\end{align}
	The result eq.\eqref{zerothpartition} matches with the known results \cite{Iliesiu:2020zld}. 	
	
The path integral with an appropriate choice of the contour of integration for the metric and dilaton thus serves as a boundary condition that chooses a particular value for $\rho(M)$ thereby specifying the state uniquely.

\section{On-shell contribution to the wave function}
\label{wdwcent}
We are interested in evaluating the wave function for the Centaur geometry. For that purpose, as mentioned above, we need input from the path integral in the form of the boundary condition to fix the function $\rho(M)$ in eq.~\eqref{psioncdil}. 
Evaluating the path integral exactly for the Centaur geometries is a  non-trivial calculation. In this paper, we settle for the modest goal of approximating the exact wave function by its saddle point value. This is generally true in the WKB approximation. In the context of 2D JT gravity for a general potential, this corresponds to the limit where 
\begin{align}
	l\gg 1 \,, \quad W(\phi_b)\gg 1 \,, \quad \frac{l}{W(\phi_b)}=\text{fixed}\,, \label{wkb}
\end{align}
where $l$ is the length of the boundary {{and $\phi_b$ is the boundary value of the dilaton }}

We shall obtain the density of states by evaluating the on-shell action for a general potential, both analytically and numerically. We outline the analytical form of the density of states in the WKB limit in two ways. First, we can work in the micro-canonical ensemble, at a fixed energy, and compute the density of states as the exponential of the entropy of the saddle point corresponding to the dominant horizon. Alternatively, working in the canonical ensemble, at a fixed temperature, we can evaluate the density of states as the inverse Laplace transform of the finite temperature partition function. We shall evaluate in both ways and also numerically compute the inverse Laplace transform and compare the results. 

The classical solutions were already discussed in section \ref{rev}. It is straightforward to compute the on-shell action from here. Consider the general solution as written in eq.~\eqref{onsol}. The function $f(r)$ can be written as
\begin{align}
	f(r)= W(r)- 2 E = W(r)-  M   \label{frm}
\end{align} 
where 
\begin{align}
	W(r)=\int_0^r U(\phi)d\phi, \quad
	 M =\int_0^{r_h}U(\phi)d\phi\label{wrm}
\end{align}
The parameter $M$ is related to the mass of the black hole $E$ as \eqref{defofM}.  For a general potential, to obtain $\rho(M)$, we need to compute the horizon value of the dilaton in terms of the mass parameter $M$ appearing in the metric as eq.~\eqref{frm}. This is usually difficult for a general potential.

For a solution which asymptotes to AdS at the boundary, the value of the dilaton at the horizon can be obtained by solving the equation 
\begin{align}
	f(\phi_h)=W(\phi_h)-   M  =0\label{phihfore1}
\end{align}
The density of states is then given {by exponentiating the entropy\footnote{ The importance of this point was emphasized to us by Sandip P. Trivedi.}},
\begin{align}
	\rho=e^{2\pi\phi_h}\label{dosinph1}
\end{align}
Another way to see this yields the correct density of states at leading order is by computing the on-shell action at finite temperature. Working in the canonical ensemble at fixed temperature is equivalent to computing the on-shell action with the renormalized length of the boundary, renormalized in units of the dilaton value at the boundary, fixed, see eq.\eqref{wkb}. 

To obtain, the on-shell action we need to evaluate the extrinsic curvature of the boundary. Taking the boundary at a constant $r$, the extrinsic curvature is found to be
\begin{align}
	K=\nabla_\mu n^\mu=\del_r\sqrt{f(r)}=\frac{U(r)}{2\sqrt{f(r)}}\label{exk}
\end{align}

The value of the on-shell action then becomes, 
\begin{align}
	I_{\text{on-shell}}&=-\half\int d{x}\int_{r_h}^{r_b} dr\left(-r U'(r)+U(r)\right)-\int_{r_b} dx \sqrt{f}r K\nonumber\\
	&=-\int d{x}\int_{r_h}^{r_b} dr \,U(r) -\half\int_{r_h} d{x} \, r U(r)\nonumber\\
	&=-\beta\left(W(r_b)-  M  +\half\phi_h U(\phi_h)\right)\label{ios1}
\end{align}
where $\beta$ is the length of the ${x}$-circle, 
{ where $M_*$ should be thought of as a function of temperature, given by 
\begin{align}
	M_*=\int_0^{r_h}U(r),\quad r_h=U^{-1}(4\pi T)\label{msdef}
\end{align}}
Using the relation in eq.~\eqref{BHtemperature} to relate $\beta$ to $U$, we find, 
\begin{align}
		I_{\text{on-shell}}=-\beta\left(W(r_b)-M_*\right)-2\pi \phi_h \label{ios2}
\end{align}
{
The term with $W(r_b)$ is divergent and hence a counterterm should be added to remove this divergence. The appropriate counter term is of the form
\begin{align}
	I_{\text{ct}}=\int_\del dx\sqrt{\gamma}\sqrt{W(\phi)}\label{Ict}
\end{align}
where $\gamma$ is the induced metric on the boundary. The on-shell action including this counterterm is given by 
\begin{align}
	I_{\text{on-shell}}\simeq \frac{\beta M_*}{2}-2\pi \phi_h\label{Iosfini}
\end{align}
The partition function is then given by 
\begin{align}
	Z=e^{-I_\text{on-shell}}=e^{-\frac{\beta M_*}{2}+2\pi\phi_h}\equiv e^{-\beta E_*+2\pi \phi_h}\label{zbeta}
\end{align}
It can be noticed from the above that the result can be thought of as the laplace transform of the function $e^{2\pi\phi_h}$ evaluated at the saddle point value for $E_*$. Thus, it follows naturally that the density of states is given by 
\begin{align}
	\rho=e^{2\pi \phi_h}\label{rhomfrozbeta}
\end{align}

}
This is the density of states corresponding to a WKB approximation of the gravity path integral. 
For a general potential, to obtain $\rho(M)$, we need to compute the horizon value of the dilaton in terms of the mass parameter $M$ appearing in the metric as eq.~\eqref{phihfore1}. This is usually difficult to write in a closed form for a general potential. So, we shall first analyze the case where the potential is a perturbation around the AdS potential i.e.,
\begin{align}
	U(\phi)=2\phi+\tilde{U}(\phi),\quad \tilde{U}(\phi)\ll\phi  \label{uinper}
\end{align}
In this case, the function $W$ defined in eq.~\eqref{wrm} can be written as
\begin{align}
	W(r)=r^2+\tilde{W}(r),\quad \tilde{W}(r)=\int^r_0 \tilde{U}(\phi)d\phi\label{winpert}
	\end{align}

The value of the dilaton at the horizon for a configuration with a given value of $M$/ can be obtained by solving the equation 
\begin{align}
	f(\phi_h)=W(\phi_h)-M&=0 \quad  
	\implies \quad \phi_h^2+\tilde{W}(\phi_h)=M\label{phihfore}
\end{align}
Solving it perturbatively in the small parameter $\tilde{W}(\phi_h)/ \phi_h^2$, we find
\begin{align}
	\phi_h \simeq \sqrt{M-\tilde{W}(\sqrt{M})} 
	\simeq\sqrt{M}-\frac{\tilde{W}(\sqrt{M})}{2\sqrt{M}}\label{phih}
\end{align}

The density of states in eq.~\eqref{rhomfrozbeta} is then given by 
\begin{align}
	\rho(M) \simeq e^{2\pi \sqrt{M}}e^{-\frac{\pi\tilde{W}(\sqrt{M})}{\sqrt{M}}} 
	\simeq e^{2\pi \sqrt{M}}\left(1-\frac{\pi\tilde{W}(\sqrt{M})}{\sqrt{M}}\right)
	\label{rhoperM}
\end{align}



\section{Explicit Wave function evaluation}
\label{explicitwf}
Finally in this section, we shall evaluate explicitly the density of states corresponding to the example mentioned in eq.~\eqref{qsimp}. 

Let us first evaluate the entropy by directly working in the microcanonical ensemble at fixed energy, related to the parameter $M$. For a fixed $M$, the horizon is given by eq.~\eqref{phihfore1}. For small enough $M$, the horizon is in the interior AdS region. The location of the horizon is given by 
\begin{align}
	W(\phi_h)=\int_0^{\phi_h} c\phi=   M \quad \Rightarrow \quad \phi_h=\sqrt{\frac{2M}{c}}\label{phihin}
\end{align}
For  large values of $M$, the horizon will be located in the AdS region connected to the asymptotic infinity. In such a case, the value of the horizon is obtained to be
\begin{align}
	& W(\phi_h)=\int_0^{\phi_1}c\phi+\int_{\phi_1}^{\phi_2}(U_0-\alpha\phi)+\int_{\phi_2}^{\phi_h}2\phi=M \nonumber\\
	&\Rightarrow \quad \half c\phi_1^2+U_0(\phi_2-\phi_1)-\half \alpha(\phi_2^2-\phi_1^2)+\phi_h^2-\phi_2^2=M\nonumber\\
	&\Rightarrow \quad \phi_h=\sqrt{M-\tilde{M}_0} \,,\quad \tilde{M}_0=\frac{(c-2)}{2}\phi_1 \phi_2\label{wphih}
\end{align}
So, we have
\begin{align}
	\rho\approx\begin{cases}
		e^{2\pi\sqrt{\frac{2M}{c}}} \, \qquad \mbox{for small $M$}\\
		e^{2\pi\sqrt{M-\tilde{M}_0}}\quad \mbox{for large $M$}
	\end{cases}
\label{rhoexpbeh}
\end{align}
Expanding the large $M$ behaviour further, we get that
\begin{align}
	\rho\xrightarrow{M\gg 1}e^{2\pi\sqrt{M}}\left(1-\frac{\pi \tilde{M}_0}{\sqrt{M}}\right)\label{rholargm}
\end{align}
which is of the form eq.~\eqref{rhoperM} with the identification $\tilde{W}(\sqrt{M})=\tilde{M}_0$.

\begin{figure}
	\centering
	\includegraphics[scale=0.7]{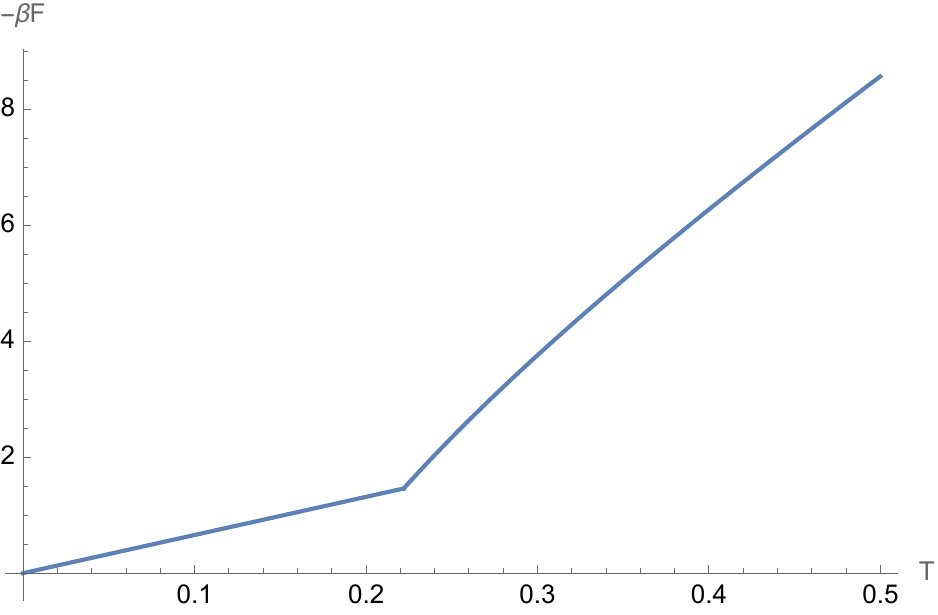}
	\caption{$- \beta F = \log Z$ as a function of $T$. The derivative of the free energy jumps at $\beta = \beta_c$ due to the first-order phase transition.}
		\label{logzt}
\end{figure}

Let us now outline an alternative way of obtaining the same result by working in the canonical ensemble. 
Working in the canonical ensemble at a fixed temperature, we can evaluate the entropy and free energy as a function of temperature. 
The location of the horizon is obtained by solving for $\phi_h$ from eq.~\eqref{BHtemperature}, 
\begin{align}
	\phi_h=\begin{cases}
	\frac{4\pi T}{c}&\quad \phi_h<\phi_1\\
	2\pi T&\quad \phi_h>\phi_2 \label{phihca}
	\end{cases}
\end{align}
For the horizon such that $\phi_h<\phi_1$ and $U(\phi_h)>U(\phi_2)$, we have the following for the entropy, energy and free energy as
\begin{align}
	S&=2\pi \phi_h=\frac{8\pi^2 T}{c}\nonumber\\
	E&=\frac{4\pi^2T^2}{c}\nonumber\\
	F&=-\frac{4\pi^2T^2}{c}\label{phlph1}
\end{align}
For the horizon such that $\phi_h>\phi_2$ and $U(\phi_h)<U(\phi_1)$, we have
\begin{align}
	S&={4\pi^2T}\nonumber\\
	E&=2\pi^2 T^2+2\pi^2\left(1-\frac{2}{c}\right)T_c^2\nonumber\\
	F&=-2\pi^2T^2+2\pi^2\left(1-\frac{2}{c}\right)T_c^2
	\label{phgph2}
\end{align}
where $T_c$ is the critical temperature which delineates the two behaviours above and is given in terms of the parameters $c,\phi_1,\phi_2$ as
\begin{align}
	T_c=\frac{1}{2 \pi}\sqrt{\frac{c \phi_1 \phi_2}{2}}\,\,\label{Tc}
\end{align}
Expressing the entropy in terms of the energy or equivalently in terms of the parameter $M$, related to the energy $E$ as $E=\frac{M}{2}$, we get that 
\begin{align}
	S=\begin{cases}
		2\pi\sqrt{\frac{2M}{c}}&\quad T<T_c\\
		2\pi \sqrt{M-4\pi^2\left(1-\frac{2}{c}\right)T_c^2}&\quad T>T_c
	\end{cases}
\label{sine}
\end{align}
It can be easily checked that the free energy is continuous at $T=T_c$, however, its derivative is not, see Fig.~\ref{logzt}. 
So, in the canonical ensemble, there is a phase transition at $T=T_c$ as one varies the temperature. The form of the entropy, and hence the corresponding density of states obtained above, matches with that of eq.~\eqref{rhoexpbeh}, as it should. In particular, we see that $\tilde{M}_0$ in eq.~\eqref{wphih} that appears as the shift in $M$ is the same that appears in eq.~\eqref{sine}, rather expressed in terms of $T_c$. We have
\begin{align}
	\tilde{M}_0=4\pi^2\left(1-\frac{2}{c}\right)T_c^2\label{Mtc}
\end{align}

The partition function at finite temperature, obtained by evaluating the on-shell action, is given by 
\begin{align}
	Z(\beta)&=\theta (\beta-\beta_c)e^{-\beta F_1}+\theta (\beta_c-\beta)e^{-\beta F_2}\nonumber\\
	F_1&=-\frac{4\pi^2}{c\beta^2}\,\,\nonumber\\
	F_2&=-\frac{2\pi^2}{\beta^2}+\frac{2\pi^2}{\beta_c^2}\left(1-\frac{2}{c}\right)\label{partinthe}
\end{align}

Let us end with a remark. One important point to note is that we have only computed the on-shell action and not the one loop determinant around the saddle point. For pure AdS, this gives $T^\frac{3}{2}$. For the case of dS bubble embedded inside AdS, the computation of this one loop factor is difficult. However, the density of states obtained from the classical analysis already shows that the number of degrees of freedom are decreased due to the presence of IR dS bubble. 

	\section{Conclusions}
	\label{concl}
	
	In this paper, we studied the canonical quantization of JT gravity with a general potential, following the approach of \cite{Hennauxjt}. Our interest and emphasis was on potentials which give rise to a geometry which is asymptotically AdS but is not pure AdS in the interior. We call such mixed geometries as Centaur geometries. Given the poor understanding of holography in dS spacetime, we focussed on potentials which gave rise to asymptotically AdS geometry but is dS spacetime in the interior. 
	We began by studying the classical solutions in JT gravity corresponding to these potentials. 
	We computed the density of states for the Centaur geometries by evaluating the on-shell value of the action. The density of states for this geometry is found to be {\it smaller} than that of the pure AdS case. This is a bit surprising as it implies that the modification of the pure AdS geometry to contain a dS bubble decreases the degrees of freedom in the spacetime. Naively, one might expect that the modification of the IR geometry, which can be regarded as an excitation of the empty AdS spacetime, should lead to an increase in the density of states. However, at the leading WKB level, we found that the opposite happens. 
	
	In this paper, we were only able to evaluate the path integral in the saddle point limit. However, it would be very interesting to compute the one-loop determinant around the saddle point we computed, in the asymptotic limit where the dilaton and the length af the boundary diverge with some combination of their ratio held fixed. This would then lead to a more complete path integral which would also mean a more exact density of states. For example, the path integral for the potential in eq.~\eqref{qsimp}, one may consider evaluating it by schematically splitting is as
	\begin{align}
		Z\sim\int \mathcal{D}g\int d\phi_1d\phi_2\int_{\phi=-\infty}^{\phi=\phi_1}\mathcal{D}\phi e^{-\int \phi(R+c)}\int_{\phi=\phi_1}^{\phi=\phi_2}\mathcal{D}\phi e^{-\int \phi(R-\alpha)+U_0}\int_{\phi=\phi_2}^{\phi=\infty}\mathcal{D}\phi e^{-\int \phi(R+2)}\label{zsp}
	\end{align}
	The evaluation of the each path integral over $\phi$ have to be done with an appropriate choice of contour.
	We expect that such a more complete density of states would pave way for a better understanding of the dS holography. Another important question would be to understand the nature of $T\bar{T}$ deformation in our geometry. 	It would be interesting to compare the nature of $T\bar{T}$ deformed geometries with the ones studied in the literature \cite{Coleman:2021nor}.

\acknowledgments
	\label{ack}
	The work of NI was supported in part by JSPS KAKENHI Grant Number 18K03619. The work of NI and SS was also by MEXT KAKENHI Grant-in-Aid for Transformative Research Areas A ``Extreme Universe'' No.~21H05184. 
	We thank Nicolo Zenoni for the helpful discussions. 
	The work of NI and SS were also supported by MEXT KAKENHI Grant-in-Aid for Transformative Research Areas A “Extreme Universe” No. 21H05184. 
	
	\bibliographystyle{JHEP}
	\bibliography{refs}

\end{document}